\documentclass[12pt,preprint]{aastex}

\def\dd{{\rm d}}

\shorttitle{Linear analysis of P-J instability with CRs}
\shortauthors{Kuwabara and Ko}

\begin{document}


\title{PARKER-JEANS INSTABILITY OF GASEOUS DISKS INCLUDING THE EFFECT OF COSMIC RAYS}


\author{Takuhito Kuwabara}
\affil{Applied Research and Standards Department,
       National institute of Information and Communications Technology,
       Koganei, Tokyo 184-8795, Japan}
\email{kuwabrtk@cc.nao.ac.jp}

\and

\author{Chung-Ming Ko}
\affil{Department of Physics, Institute of Astronomy and Center
       for Complex Systems, National Central University,
       Chung-Li, Taiwan 320, Republic of China}
\email{cmko@astro.ncu.edu.tw}



\begin{abstract}
We use linear analysis to examine the effect of cosmic rays (CRs) on the Parker-Jeans
instability of magnetized self-gravitating gaseous disks.
We adopt a slab equilibrium model in which the gravity (including self-gravity) is
perpendicular to the mid-plane, the magnetic field lies along the slab.
CR is described as a fluid and only along magnetic field lines diffusion is considered.
The linearised equations are solved numerically.
The system is susceptible to Parker-Jeans instability.
In general the system is less unstable when the CR diffusion coefficient is smaller
(i.e., the coupling between the CRs and plasma is stronger).
The system is also less unstable if CR pressure is larger.
This is a reminiscence of the fact that Jeans instability and Parker instability are less
unstable when the gas pressure is larger (or temperature is higher).
Moreover, for large CR diffusion coefficient (or small CR pressure),
perturbations parallel to the magnetic field are more unstable than
those perpendicular to it.
The other governing factor on the growth rate of the perturbations in different directions
is the thickness of the disk or the strength of the external pressure on the disk.
In fact, this is the determining factor in some parameter regimes.
\end{abstract}




\section{INTRODUCTION}
In our Galaxy (and probably other galaxies as well), the energy densities of
thermal gas, magnetic field, cosmic rays and turbulence are of the same order
of magnitude
\citep[$\sim$1 eV cm$^{-3}$, see e.g.,][]{gaisser90}.
Rough equipartition exists between these components indicates that they are
tightly coupled together, and all of them should have significant influence
on the structure and evolution of the interstellar medium (ISM).
However, the dynamical role of CRs on ISM has not been taken seriously in most
ISM related models until rather recently
\citep[see e.g.,][]{ferriere01}.

The ISM is highly inhomogeneous and have many different phases of gas clouds,
e.g., hot phase, warm phase, diffuse clouds, molecular clouds, etc.
This is due partly to the existence of many instabilities in ISM conditions.
Gravitational instability (Jeans instability) and magnetic buoyancy instability
\citep[Parker instability,][]{parker66}
have been considered to be responsible for the formation of some large scale clouds
(e.g., HI clouds, molecular clouds).
As ISM is aggregated, filament-like structures are formed by Parker-Jeans instability.
Extensive analytical works on Parker-Jeans instability have been carried out in the
past two to three decades
\citep[e.g.,][]{elmegreen78,elmegreen82a,elmegreen82b,hanawa92,nagai98,lee05}.
Recently, this instability has been studied by the magnetohydrodynamic (MHD) simulations
\citep[e.g.,][]{chou00,lee01,kim02}.
\citet*{nagai98} showed that the angle of the filament formed by this
instability with respect to the magnetic field lines depends on the
strength of the external gas pressure of the initial equilibrium gas layer, and
\citet*{chou00} confirmed their results by MHD simulations.
Nevertheless, the dynamical effect of CRs has not been addressed in these works.

In his original work on the Parker instability, Parker realised the importance of CRs
on the instability and did consider their effect in a simplified way
\citep{parker66}.
However, studies on the instability with more rigorous treatment of the dynamics of
CRs appear only rather recently.
Both linear analysis
\citep{kim97,kim98,hanasz97I,hanasz97II,ryu03}
and MHD simulation
\citep{hanasz00,hanasz03,hanasz04I,hanasz04II,kuwabara04}
have been performed on models in which the diffusion of CRs through plasma is included.
These models are based on the hydrodynamic description of CRs
\citep[e.g.,][]{drury81,drury83,ko92},
which is a fairly good approximation in studying the structure and evolution of the system,
although the spectrum of CR is compromised.
As far as we know, the effect of the CR diffusion on the Parker-Jeans instability
has not been studied yet.

In this paper, the effect of CRs on the Parker-Jeans instability is examined by
linear analysis. Hydrodynamic approach to CR is adopted and only along field line diffusion
is considered. As a model for Galactic gaseous disk or compressed molecular cloud,
a self-gravitating gas layer or slab threaded by magnetic field perpendicular
to the gravity is used for the unperturbed equilibrium state.
The dispersion relations are obtained numerically.
The paper is organised as follows.
In \S\,2, we present the CR-plasma system and the equilibrium model.
\S\,3 is devoted to linear stability analysis, and
the results are described in \S\,4.
\S\,5 provides a summary and discussion.

\section{MODELS}

\subsection{Two-Fluid Model}
We adopt a two-fluid model which comprises cosmic rays and thermal gas
\citep[e.g.,][]{drury81,drury83,ko92}.
CR is treated as a massless gas with significant energy density or pressure.
CRs are coupled to the thermal gas via hydromagnetic irregularities or waves embedded
in the flow. To a first approximation, the only effect of the irregularities or waves
is contained in the diffusion coefficient of CR.
For simplicity, we consider diffusion along magnetic field lines only
and ignore the cross-field-line diffusion
\citep[see e.g.,][]{giacalone99}.
Thus ${\bf\kappa}=\kappa_\|{\bf b}{\bf b}$, where ${\bf b}$ is the unit vector along
the magnetic field.

Assuming infinite conductivity for the plasma (i.e., ideal MHD),
the system is governed by the total mass equation, momentum equation, thermal and
CR energy equations,
\begin{equation}\label{mass}
  {\partial \rho\over\partial t}
  +\nabla\cdot\left(\rho{\bf V}\right)=0\,,
\end{equation}
\begin{eqnarray}\label{momentum}
  {\partial\over\partial t}\left(\rho{\bf V}\right)
  +\nabla\cdot\left[\rho{\bf V}{\bf V}
  +\left(P_{\rm g}+P_{\rm c}+{B^2\over 2\mu_0}\right){\bf I}
  -{{\bf B}{\bf B}\over \mu_0}\right] \nonumber \\
  \quad\quad +\rho\left[\nabla(\psi+\psi_{\rm e})+2{\bf \Omega}\times{\bf V}
  +{\bf\Omega}\times({\bf\Omega}\times{\bf r})\right]=0\,,
\end{eqnarray}
\begin{equation}\label{gas-en}
  {\partial P_{\rm g}\over\partial t}+{\bf V}\cdot\nabla P_{\rm g}
  +\gamma_{\rm g}P_{\rm g}\nabla\cdot {\bf V}=0\,,
\end{equation}
\begin{equation}\label{cr-en}
  {\partial P_{\rm c}\over\partial t}+{\bf V}\cdot\nabla P_{\rm c}
  +\gamma_{\rm c}P_{\rm c}\nabla\cdot {\bf V}
  -\nabla\cdot\left(\kappa_\|{\bf b}{\bf b}\cdot\nabla P_{\rm c}\right)=0\,,
\end{equation}
supplemented by the Poisson equation for self-gravity and induction equation
for magnetic field
\begin{equation}\label{poisson}
  \nabla^2\psi = 4\pi G\rho\,,
\end{equation}
\begin{equation}\label{induction}
  {\partial {\bf B}\over\partial t}=\nabla\times({\bf V}\times{\bf B})\,,
\end{equation}
\begin{equation}\label{divergent}
  \nabla\cdot{\bf B}=0\,.
\end{equation}
In these equations $\rho$, ${\bf V}$, $P_{\rm g}$ and $\gamma_{\rm g}$ denote the
thermal gas density, flow velocity, pressure and polytropic index, respectively;
$P_{\rm c}$, $\gamma_{\rm c}$ and $\kappa_\|$ denote the CR pressure, polytropic
index and diffusion coefficient along magnetic field lines, respectively;
$\psi$ and $\psi_{\rm e}$ denote the potential of the self-gravity of the gas and
the potential of some external gravitational source, respectively;
${\bf B}$ is the magnetic field strength and ${\bf \Omega}$ is the angular
velocity of the disk.
In the rest of the paper we assume that the centrifugal force is balanced by the
external gravitational source.

\subsection{Equilibrium Model}
We are interested in the stability of a uniformly rotating gaseous disk or a slab.
We adopt a local Cartesian coordinate system $(x,y,z)$ such that
${\bf e_x}={\bf e_\phi}$, ${\bf e_y}=-{\bf e_r}$ and ${\bf e_z}={\bf e_z}$,
where $(r,\phi,z)$ is the cylindrical coordinate system of the disk.
All quantities of the unperturbed equilibrium state depend on $z$ only.
Moreover, assume initial magnetic field is in the azimuthal direction,
i.e., ${\bf B}=B(z) {\bf e_x}$.
For simplicity ignore the external gravitational force in the $z$-direction,
the equilibrium model satisfies the magnetohydrostatic equation
\begin{equation}\label{magnetohydrostatic}
  {\dd\over \dd z}\left(P_{\rm g}+P_{\rm c}+{B^2\over 2\mu_0}\right)
  +\rho{\dd\psi\over \dd z}=0\,.
\end{equation}
Substitute into the Poisson equation (Eq.~(\ref{poisson})), we obtain
\begin{equation}\label{equilibrium}
  {\dd\over \dd z}\left[{C_{\rm s}^2(1+\alpha+\beta)\over\gamma_{\rm g}P_{\rm t}}
  {\dd P_{\rm t}\over\dd z}\right]
  +{4\pi G\gamma_{\rm g}P_{\rm t}\over C_{\rm s}^2(1+\alpha+\beta)}=0\,,
\end{equation}
where $\alpha$ and $\beta$ are the ratio of the magnetic pressure and CR pressure
to the gas pressure, $P_{\rm t}=(1+\alpha+\beta)P_{\rm g}$ is the total pressure,
and $C_{\rm s}=(\gamma_{\rm g}P_{\rm g}/\rho)^{1/2}$ is the gas sound speed.
Moreover, we assume a temperature distribution
\begin{equation}\label{tempdist}
  T(z)=\left\{
  \begin{array}{rl}
  T_d\,,    & \quad \mbox{for $|z|<z_d$}\,, \\
  \infty\,, & \quad \mbox{for $|z|>z_d$}\,,
  \end{array}\right.
\end{equation}
\citep[][]{hanawa92},
where $T_d$ and $z_d$ are the temperature and the half-thickness
of the (cold) gaseous disk.
Large (small) $z_d$ can be interpreted as small (large) external pressure.
Integrate Eq.~(\ref{equilibrium}) with this temperature distribution gives
the equilibrium pressure and density distributions.
The density scale height at $z=0$ (mid-plane of the gaseous disk) is
$H=C_{\rm s0}\sqrt{(1+\alpha+\beta)/(2\pi G\rho_0\gamma_{\rm g})\,}$,
where the subscript $0$ denotes the value at the mid-plane.
Naturally, we normalise the density, velocity and length of the gaseous disk
to $\rho_0$, $C_{\rm s0}$ and $H_0=C_{\rm s0}/\sqrt{2\pi G\rho_0\gamma_{\rm g}\,}$.
The CR diffusion coefficient is normalised to $C_{\rm s0}H_0$.
In this paper we choose $\gamma_{\rm g}=1$ and $\gamma_{\rm c}=4/3$.
We pick $\alpha=1$, $\beta=1$ and $z_d=H_0$ as the fiducial model.
Fig.~{\ref{initial_dist}} shows the distribution of the equilibrium gas pressure
and the gravity in the case of $\alpha=1$ and $\beta=1$.

Parker-Jeans instability is one of the possible mechanisms for the formation of
large ISM clouds (e.g., molecular clouds, HI clouds), thus the typical values for
$\rho_0$, $C_{\rm s0}$, $H_0$ and $C_{\rm s0}H_0$ are
$1.67\times 10^{-21}$ kg m$^{-3}$, 5 km s$^{-1}$, 200 pc and
$3\times 10^{22}$ m$^2$ s$^{-1}$.
The effective sound speed $C_{\rm s0}\sqrt{1+\alpha+\beta\,}\sim 8.5$ km s$^{-1}$
for $\alpha=\beta=1$
\citep*[][]{hanawa92}.
The value of CR diffusion coefficient is estimated to be
$3\times 10^{24}$ m$^2$ s$^{-1}$
\citep[e.g.,][]{berezinskii90,ptuskin01,ryu03}.
Thus the fiducial value of normalised $\kappa_{\|}$ is 100.
On the other hand, Parker-Jeans instability may cause the fragmentation and collapse
of large ISM clouds. The typical values for
$\rho_0$, $C_{\rm s0}$, $H_0$ and $C_{\rm s0}H_0$ become
$1.67\times 10^{-19}$ kg m$^{-3}$, 5 km s$^{-1}$, 20 pc and
$3\times 10^{21}$ m$^2$ s$^{-1}$.
In this case the CR diffusion coefficient is smaller, say
$3\times 10^{23}$ m$^2$ s$^{-1}$, then the fiducial value of normalised $\kappa_{\|}$ is
still 100.

\section{LINEAR STABILITY ANALYSIS}

\subsection{Equations for Perturbations}
We perform standard linear stability analysis on the system of equations
Eqs.~(\ref{mass})-(\ref{divergent}) of the slab equilibrium model described in previous
section.
Since the unperturbed state depends on $z$ only, we assume the perturbed quantities to
be of the following form
\begin{equation}\label{perturbed}
\left[\begin{array}{c}
\delta\rho \\
\delta V_x \\
\delta V_y \\
\delta V_z \\
\delta P_g \\
\delta P_c \\
\delta B_x \\
\delta B_y \\
\delta B_z \\
\delta\psi \\
\end{array}\right]
=
\left[\begin{array}{c}
\delta\bar{\rho} \\
i\delta\bar{V_x} \\
i\delta\bar{V_y} \\
\delta\bar{V_z}  \\
\delta\bar{P_g}  \\
\delta\bar{P_c}  \\
\delta\bar{B_x}  \\
\delta\bar{B_y}  \\
-i\delta\bar{B_z}\\
\delta\bar{\psi} \\
\end{array}\right]\exp(\sigma t+ik_xx+ik_yy)\,,
\end{equation}
where the barred quantities are functions of $z$ only,
$\sigma$ is the growth rate, and $k_x$, $k_y$ are the wave numbers in
the $x$- and $y$-directions, respectively.
The set of linearised equations becomes
\begin{equation}\label{lin-mass}
  \sigma{\delta\bar\rho\over\rho}-k_x\delta\bar V_x-k_y\delta\bar V_y
  +{\dd\delta\bar V_z\over\dd  z}+{1\over\rho}{\dd\rho\over\dd z}\delta\bar V_z=0\,,
\end{equation}
\begin{equation}\label{lin-momentumx}
  \sigma\delta\bar V_x
  +{k_x\over\rho}\delta\bar P_{\rm g}
  +{k_x\over\rho}\delta\bar P_{\rm c}
  +{1\over\mu_0\rho}{\dd B_x\over\dd z}\delta\bar B_z
  -2\Omega\delta\bar V_y+k_x\delta\bar\psi=0\,,
\end{equation}
\begin{equation}\label{lin-momentumy}
  \sigma\delta\bar V_y+{k_y\over\rho}\delta\bar P_{\rm g}
  +{k_y\over\rho}\delta\bar P_{\rm c}-{k_xB\over\mu_0\rho}\delta\bar B_y
  +{k_yB\over\mu_0\rho}\delta\bar B_x
  +2\Omega\delta\bar V_x+k_y\delta\bar\psi=0\,,
\end{equation}
\begin{eqnarray}\label{lin-momentumz}
  \sigma\delta\bar{V_z}
  +{1\over\rho}{\dd\delta\bar P_{\rm g}\over\dd z}
  +{1\over\rho}{\dd\delta\bar P_{\rm c}\over\dd z}
  +{1\over\rho}{\dd\over\dd z}\left({B\delta\bar B_x\over\mu_0}\right)
  -{k_xB\over\mu_0\rho}\delta\bar B_z \nonumber \\
  -{1\over\rho^2}{\dd\over\dd z}\left(P_{\rm g}+P_{\rm c}+{B^2\over 2\mu_0}\right)
  +{\dd\delta\bar\psi\over\dd z}=0\,,
\end{eqnarray}
\begin{equation}\label{lin-gas-en}
  \sigma\delta\bar P_{\rm g}+{\dd P_{\rm g}\over\dd z}\delta\bar V_z
  -{\gamma_{\rm g}P_{\rm g}\over\rho}
  \left(\sigma\delta\bar\rho+{\dd\rho\over\dd z}\delta\bar V_z\right)=0\,,
\end{equation}
\begin{equation}\label{lin-cr-en}
  \sigma\delta\bar P_{\rm c}+{\dd P_{\rm c}\over\dd z}\delta\bar V_z
  -{\gamma_{\rm c}P_{\rm c}\over\rho}
  \left(\sigma\delta\bar\rho+{\dd\rho\over\dd z}\delta\bar V_z\right)
  +\kappa_\|k_x^2\delta\bar P_{\rm c}
  -{\kappa_\|k_x\over B}{\dd P_{\rm c}\over\dd z}\delta\bar B_z=0\,,
\end{equation}
\begin{equation}\label{lin-poisson}
  -(k_x^2+k_y^2)\delta\bar\psi+{\dd^2\delta\bar\psi\over\dd z^2}
  -4\pi G\delta\bar\rho=0\,,
\end{equation}
\begin{equation}\label{lin-inductionx}
  \sigma\delta\bar B_x+k_xB\delta\bar V_x
  +{\dd B\over\dd z}\delta\bar V_z
  -{B\over\rho}
  \left(\sigma\delta\bar\rho+{\dd\rho\over\dd z}\delta\bar V_z\right)
  =0\,,
\end{equation}
\begin{equation}\label{lin-inductiony}
  \sigma\delta\bar B_y+k_xB\delta\bar V_y=0\,,
\end{equation}
\begin{equation}\label{lin-inductionz}
  \sigma\delta\bar B_z+k_xB\delta\bar V_z=0\,,
\end{equation}
\begin{equation}\label{lin-divergent}
  k_x\delta\bar B_x+k_y\delta\bar B_y-{\dd\delta\bar B_z\over\dd z}=0\,.
\end{equation}
After some manipulations, we obtain the following set of ordinary differential equations,
\begin{equation}\label{odes}
  {\dd\over\dd z}\left[\begin{array}{c}
             y_1 \\
             y_2 \\
             y_3 \\
             y_4
           \end{array}\right]
=
\left[\begin{array}{cccc}
      A_{11} & A_{12} & A_{13} & 0      \\
      A_{21} & A_{22} & A_{23} & A_{24} \\
      0      & 0      & 0      & -1     \\
      A_{41} & A_{42} & A_{43} & 0
\end{array}\right]\cdot
\left[\begin{array}{c}
      y_1 \\
      y_2 \\
      y_3 \\
      y_4
      \end{array}\right]\,,
\end{equation}
where the running variables are defined as
\begin{equation}\label{y1}
  y_1\equiv -{\rho\delta\bar V_z\over\sigma}\,,
\end{equation}
\begin{equation}\label{y2}
  y_2\equiv\delta\bar P_{\rm t}
  =\delta\bar P_{\rm g}+\delta\bar P_{\rm c}+{B\over\mu_0}\delta\bar B_x\,,
\end{equation}
\begin{equation}\label{y3}
  y_3\equiv\delta\bar\psi\,,
\end{equation}
\begin{equation}\label{y4}
  y_4\equiv\delta\bar g_z=-{\dd\delta\bar\psi\over\dd z}\,,
\end{equation}
and the matrix elements are
\begin{eqnarray}\label{a11}
  A_{11}=
  \left[1-{\sigma^2\over\Sigma^2}
       -{V_{\rm A}^2k_x^2\over\Sigma^2(1+\Gamma^2)}
       \left(\Gamma^2+{k_y\over k_x}\Gamma\right)\right]
  {\dd\ln\rho\over\dd z}
  \qquad\qquad\qquad\qquad\qquad\qquad \nonumber \\
  -{1\over D}
  \left[{\sigma^2\over\Sigma^2}
       +{V_{\rm A}^2k_x^2\over\Sigma^2(1+\Gamma^2)}
       \left(\Gamma^2+{k_y\over k_x}\Gamma\right)\right]
  \qquad\qquad\qquad\qquad\qquad\qquad\qquad \nonumber \\
  \times
  \left\{V_{\rm A}^2{\dd\ln B\over\dd z}+{1\over\rho}{\dd P_{\rm c}\over\dd z}
       -\left\{{C_{\rm c}^2\over(1+K)}
       +V_A^2\left[{\sigma^2\over\Sigma^2}
       +{k_x^2V_{\rm A}^2\Gamma^2\over\Sigma^2(1+\Gamma^2)}\right]\right\}
  {\dd\ln\rho\over\dd z}\right\}\,,
\end{eqnarray}
\begin{eqnarray}\label{a12}
  A_{12}={(k_x^2+k_y^2)\over\Sigma^2(1+\Gamma^2)}
  +{1\over D}
  \left[{\sigma^2\over\Sigma^2}
       +{V_{\rm A}^2k_x^2\over\Sigma^2(1+\Gamma^2)}
       \left(\Gamma^2+{k_y\over k_x}\Gamma\right)\right] \nonumber \\
  \times
  \left[{\sigma^2\over\Sigma^2}
       +{V_{\rm A}^2k_x^2\over\Sigma^2(1+\Gamma^2)}
       \left(\Gamma^2-{k_y\over k_x}\Gamma\right)\right]\,,
\end{eqnarray}
\begin{equation}\label{a13}
  A_{13}={(k_x^2+k_y^2)\rho\over\Sigma^2(1+\Gamma^2)}
  -{1\over D}{V_{\rm A}^2k_x^2\rho\over\Sigma^2(1+\Gamma^2)}
  \left(1+{k_y\over k_x}\Gamma\right)
  \left[{\sigma^2\over\Sigma^2}
       +{V_{\rm A}^2k_x^2\over\Sigma^2(1+\Gamma^2)}
       \left(\Gamma^2+{k_y\over k_x}\Gamma\right)\right]\,,
\end{equation}
\begin{eqnarray}\label{a21}
  A_{21}=
  \sigma^2+V_{\rm A}^2k_x^2
  \qquad\qquad\qquad\qquad\qquad\qquad\qquad\qquad
  \qquad\qquad\qquad\qquad\qquad \nonumber \\
  -{g_z\over D}
  \left\{V_{\rm A}^2{\dd\ln B\over\dd z}+{1\over\rho}{\dd P_{\rm c}\over\dd z}
       -\left\{{C_{\rm c}^2\over(1+K)}
       +V_A^2\left[{\sigma^2\over\Sigma^2}
       +{k_x^2V_{\rm A}^2\Gamma^2\over\Sigma^2(1+\Gamma^2)}\right]\right\}
  {\dd\ln\rho\over\dd z}\right\}\,,
\end{eqnarray}
\begin{equation}\label{a22}
  A_{22}=
  {g_z\over D}
  \left[{\sigma^2\over\Sigma^2}
       +{V_{\rm A}^2k_x^2\over\Sigma^2(1+\Gamma^2)}
       \left(\Gamma^2-{k_y\over k_x}\Gamma\right)\right]\,,
\end{equation}
\begin{equation}\label{a23}
  A_{23}=-{g_z\over D}{V_{\rm A}^2k_x^2\rho\over\Sigma^2(1+\Gamma^2)}
  \left(1+{k_y\over k_x}\Gamma\right)\,,
\end{equation}
\begin{equation}\label{a24}
  A_{24}=\rho\,,
\end{equation}
\begin{equation}\label{a41}
  A_{41}={4\pi G\over D}
  \left\{V_{\rm A}^2{\dd\ln B\over\dd z}+{1\over\rho}{\dd P_{\rm c}\over\dd z}
       -\left\{{C_{\rm c}^2\over(1+K)}
       +V_A^2\left[{\sigma^2\over\Sigma^2}
       +{k_x^2V_{\rm A}^2\Gamma^2\over\Sigma^2(1+\Gamma^2)}\right]\right\}
  {\dd\ln\rho\over\dd z}\right\}\,,
\end{equation}
\begin{equation}\label{a42}
  A_{42}=-{4\pi G\over D}
  \left[{\sigma^2\over\Sigma^2}
       +{V_{\rm A}^2k_x^2\over\Sigma^2(1+\Gamma^2)}
       \left(\Gamma^2-{k_y\over k_x}\Gamma\right)\right]\,,
\end{equation}
\begin{equation}\label{a43}
  A_{43}=-(k_x^2+k_y^2)
  +{4\pi G\over D}{V_{\rm A}^2k_x^2\rho\over\Sigma^2(1+\Gamma^2)}
  \left(1+{k_y\over k_x}\Gamma\right)\,,
\end{equation}
in which
\begin{eqnarray}\label{variables1}
  K={\kappa_\|k_x^2\over\sigma}\,,
  \quad \Sigma^2=\sigma^2+V_{\rm A}^2k_x^2\,,
  \quad \Gamma={2\sigma\Omega\over\Sigma^2}\,, \nonumber \\
  \quad D=C_{\rm s}^2+{C_{\rm c}^2\over(1+K)}
  +V_A^2\left[{\sigma^2\over\Sigma^2}
  +{V_A^2k_x^2\Gamma^2\over\Sigma^2(1+\Gamma^2)}\right]\,,
\end{eqnarray}
\begin{equation}\label{variables2}
  C_{\rm s}^2={\gamma_{\rm g}P_{\rm g}\over\rho}\,,
  \quad C_{\rm c}^2={\gamma_{\rm c}P_{\rm c}\over\rho}\,,
  \quad V_A^2={B^2\over\mu_0\rho}\,,
  \quad g_z={1\over\rho}{\dd P_{\rm t}\over\dd z}\,.
\end{equation}
For completeness, we list the perturbed quantities in terms of the
running variables (\ref{y1})-(\ref{y4}),
\begin{eqnarray}\label{drho}
  \delta\bar\rho=
  -\left\{V_{\rm A}^2{\dd\ln B\over\dd z}+{1\over\rho}{\dd P_{\rm c}\over\dd z}
       -\left\{{C_{\rm c}^2\over(1+K)}
       +V_A^2\left[{\sigma^2\over\Sigma^2}
       +{k_x^2V_{\rm A}^2\Gamma^2\over\Sigma^2(1+\Gamma^2)}\right]\right\}
       {\dd\ln\rho\over\dd z}\right\}{y_1\over D} \nonumber \\
  +\left[{\sigma^2\over\Sigma^2}
       +{V_{\rm A}^2k_x^2\over\Sigma^2(1+\Gamma^2)}
       \left(\Gamma^2-{k_y\over k_x}\Gamma\right)\right]{y_2\over D}
  -{V_{\rm A}^2k_x^2\rho\over\Sigma^2(1+\Gamma^2)}
       \left(1+{k_y\over k_x}\Gamma\right){y_3\over D}\,,
\end{eqnarray}
\begin{equation}\label{dvelx}
  \delta\bar V_x=
  -{\sigma k_x\over\Sigma^2(1+\Gamma^2)}
  \left[\left(1+{k_y\over k_x}\right)\left(y_3+{y_2\over\rho}\right)
       +V_{\rm A}^2\left({1\over\rho^2}{\dd\rho\over\dd z}y_1
       -{\delta\bar\rho\over\rho}\right)\right]\,,
\end{equation}
\begin{equation}\label{dvely}
  \delta\bar V_y=
  -{\sigma k_y\over\Sigma^2(1+\Gamma^2)}
  \left[\left(1+{k_x\over k_y}\right)\left(y_3+{y_2\over\rho}\right)
       +V_{\rm A}^2\Gamma{k_x\over k_y}
       \left({1\over\rho^2}{\dd\rho\over\dd z}y_1
       -{\delta\bar\rho\over\rho}\right)\right]\,,
\end{equation}
\begin{equation}\label{dvelz}
  \delta\bar V_z=-{\sigma\over\rho}y_1\,,
\end{equation}
\begin{equation}\label{dpg}
  \delta\bar P_{\rm g}=C_{\rm s}^2\delta\bar\rho
  -{1\over\rho}\left(C_{\rm s}^2{\dd\rho\over\dd z}-{\dd P_{\rm g}\over\dd z}
  \right)y_1\,,
\end{equation}
\begin{equation}\label{dpc}
  \delta\bar P_{\rm c}={C_{\rm c}^2\over(1+K)}\delta\bar\rho
  -{1\over\rho}\left[{C_{\rm c}^2\over(1+K)}{\dd\rho\over\dd z}
  -{\dd P_{\rm c}\over\dd z}\right]y_1\,,
\end{equation}
\begin{equation}\label{dpsi}
  \delta\bar\psi=y_3\,,
  \quad\quad\delta\bar g_z=y_4\,,
\end{equation}
\begin{equation}\label{dbx}
  \delta\bar B_x=
  {B\over\rho}\left({\dd\ln B\over\dd z}-{\dd\ln\rho\over\dd z}\right)y_1
  +B{\delta\bar\rho\over\rho}-{k_x\over\sigma}B\delta\bar V_x\,,
\end{equation}
\begin{equation}\label{dby}
  \delta\bar B_y=-{k_x\over\sigma}B\delta\bar V_y\,,
\end{equation}
\begin{equation}\label{dbz}
  \delta\bar B_z=-{k_x\over\sigma}B_x\delta\bar V_z\,,
\end{equation}

By setting up suitable boundary conditions at $z=0$ and $z=z_d$,
the set of equations (Eq. (\ref{odes})) is solved for given $(k_x,k_y)$.
The growth rate is the corresponding eigenvalue that satisfies the
boundary conditions.

\subsection{Boundary Conditions}
The slab geometry and layered equilibrium model (Eq.~(\ref{tempdist}))
we use in \S\,2 is a kind of disk-halo model.
We adopt the boundary conditions used by \citet*{nagai98}, \citet*{hanawa92}
and \citet*{chou00}
for the interface between disk and halo (at $z=z_d$).
The boundary conditions are derived from the continuity of
velocity, pressure, magnetic field, gravitational potential,
and gravitational acceleration across the interface $z=z_d$,
\begin{equation}
  y_2(z_d)=\left[g_z(z_d)-{k_x^2\over\sqrt{k_x^2+k_y^2\,}}\,V_{\rm A}^2
  \right]y_1(z_d)\,,
\end{equation}
and
\begin{equation}\label{bc-zd}
  y_4(z_d)=-4\pi Gy_1(z_d)+\sqrt{k_x^2+k_y^2\,}\,y_3(z_d)\,.
\end{equation}

The natural boundary conditions at the mid-plane ($z=0$) are,
\begin{equation}\label{bc-0}
  \left\{
   \begin{array}{cl}
   y_1(0)=y_4(0)=0\,, & \mbox{ (in the case of symmetry)}\,, \\
   y_2(0)=y_3(0)=0\,, & \mbox{ (in the case of antisymmetry)}\,.
   \end{array}\right.
\end{equation}




\section{RESULTS}
In this section we present the dispersion relation of the Parker-Jeans instability
with respect to different parameters of the system, such as
the CR diffusion coefficient $\kappa_{\|}$,
the initial ratio of the CR pressure to gas pressure $\beta$,
the initial half-thickness of the gaseous disk $z_d$, and
the angular velocity $\Omega$.
We choose $\alpha=1$, $\beta=1$, $z_d/H_0=1$, $\kappa_\|/C_{\rm s0}H_0=100$, $\Omega=0$
as the fiducial model (see \S\,2 for the normalisations).
The values of $z_d$ and $\kappa_\|$ in the following discussion are the
normalised values.

We only present the results of symmetric boundary at $z=0$,
because the growth rate in the case of antisymmetric boundary at $z=0$
is much smaller
\citep*[e.g.,][]{chou00}.

In the following, $\sigma_x$ and $\sigma_{x,{\rm max}}$ denote the growth rate
and maximum growth rate when $k_y=0$, i.e., no perturbations perpendicular to
the unperturbed magnetic field; and $\sigma_y$ and $\sigma_{y,{\rm max}}$
denote the growth rate and maximum growth rate when $k_x=0$,
i.e., no perturbations parallel to the unperturbed magnetic field.
$k_{x,{\rm crit}}$ ($k_{y,{\rm crit}}$) denotes the critical wavenumber
beyond which the system becomes stable (i.e., $\sigma_x<0$) when $k_y=0$ ($k_x=0$).
We are particular interested in these two perturbations ($k_y=0$ and $k_x=0$),
because from the numerically studies we have so far, the most unstable mode of the
system always occurs at either $k_y=0$ or $k_x=0$.
That means the magnetised gaseous disk tends to break up into elongated fragments
with axes either perpendicular to ($k_y=0$) or parallel to ($k_x=0$) the initial
magnetic field.

\subsection{Dependence on CR Diffusion Coefficient $\kappa_{\|}$}
Fig.~\ref{sig_x_k} ({\it top}) shows the dispersion relation of the case $k_y=0$
for different $\kappa_\|$ (the CR diffusion coefficient).
The other parameters take the fiducial values $\alpha=1$, $\beta=1$, $z_d=1$
and $\Omega=0$.
As $\kappa_\|$ increases, the growth rate $\sigma_x$ becomes larger and
the short wavelength perturbations become more and more unstable.
The most unstable mode (the wavenumber at which $\sigma_x$ attains its
maximum value) increases as $\kappa_\|$ increases.
We note that $k_{x,{\rm crit}}$, the critical wavenumber beyond which the system
becomes stable is different for $\kappa_\|=0$ and $\kappa_\|\ne 0$.
This remind us about the effect of diffusion of CR on Jeans instability
of uniform unperturbed state.
When CR and gas are tightly bound together, the stability criterion depends
on the sum of the gas and CR pressures.
If there is CR diffusion (no matter how small it is), the stability criterion
depends on the gas pressure only.
The only effect of CR is on the growth rate and not the stability criterion.

Fig.~\ref{sig_x_k} ({\it bottom}) shows the dependence of the maximum
growth rate $\sigma_{x,{\rm max}}$ on $\kappa_{\|}$.
$\sigma_{x,\rm max}$ ranges from $\sim 0.53$ to $\sim 0.65$
in the range of $0<\kappa_\|<200$.
When $\kappa_\|<10$, $\sigma_{x,{\rm max}}$ increases rapidly with $\kappa_\|$,
then it levels off to almost a constant when $\kappa_\|>100$.

Fig.~\ref{sig_y_k} shows the dispersion relation of the case $k_x=0$
for different $\kappa_\|$.
The other parameters take the fiducial values $\alpha=1$, $\beta=1$, $z_d=1$
and $\Omega=0$.
$\sigma_y$ does not depend on $\kappa_\|$ at all,
which is understandable as CRs diffuse along magnetic field lines only.

Fig.~\ref{sig_xy_k} shows the dispersion relation of general perturbations
for $\kappa_\|=100$ ({\it left}), $1$ ({\it middle}), and $0.1$ ({\it right}).
The other parameters take the fiducial values $\alpha=1$, $\beta=1$, $z_d=1$
and $\Omega=0$.
Since $\sigma_{x,{\rm max}}$ decreases as $\kappa_\|$ decreases while
$\sigma_{y,{\rm max}}$ is independent of $\kappa_\|$, it is expected that the
most unstable mode shifts from $k_y=0$ to $k_x=0$ as $\kappa_\|$ decreases.
In fact, the transition occurs somewhere around $\kappa_\|\sim 1$
(Fig.~\ref{sig_xy_k} {\it middle}).
As the coupling between CR and gas increases, the short wavelength perturbations
in the unperturbed magnetic field direction are inhibited.

Although the system is stable for large $k_y$ when $k_x=0$,
it is unstable when $k_x$ is small but not zero.
Moreover, the growth rate at large $k_y$ and small $k_x$ becomes smaller
as $\kappa_\|$ decreases.

\subsection{Dependence on Ratio of CR Pressure to Gas Pressure $\beta$}
Fig.~\ref{betax} shows the dispersion relation of the case $k_y=0$
for different $\beta$ (the ratio of the unperturbed CR pressure to gas pressure).
The other parameters take the fiducial values $\alpha=1$, $\kappa_\|=100$,
$z_d=1$ and $\Omega=0$.
$\sigma_{x,{\rm max}}$ increases (decreases) with $\beta$ when $\beta<3$ ($>3$).
The critical wavenumber $k_{x,{\rm crit}}$ increases with $\beta$ for $\beta<3$,
and it remains roughly the same for $\beta>3$.
As mentioned in previous subsection that CR should not affect the Jeans stability
criterion of uniform unperturbed state when the diffusion coefficient is nonzero.
The dependence on $\beta$ of the critical wavenumber in this case comes from the
dependence of the scale height of the unperturbed nonuniform state on the
unperturbed CR pressure.

Fig.~\ref{betay} shows the dispersion relation of the case $k_x=0$
for different $\beta$.
The other parameters take the fiducial values $\alpha=1$, $\kappa_\|=100$,
$z_d=1$ and $\Omega=0$.
$\sigma_{y,{\rm max}}$ decreases as $\beta$ increases and levels off at $\beta>10$.

Fig.~\ref{w3beta} shows the dispersion relation of general perturbations
for $\beta=1$ ({\it left}), $40$ ({\it middle}), and $100$ ({\it right}).
The other parameters take the fiducial values $\alpha=1$, $\kappa_\|=100$,
$z_d=1$ and $\Omega=0$.
At small $\beta$ the most unstable mode has $k_y=0$.
Around $\beta\sim 40$, $\sigma_{x,{\rm max}}\approx\sigma_{y,{\rm max}}$ and
the most unstable mode shifts to $k_x=0$ for larger $\beta$.
Moreover, the growth rate at large $k_y$ and small $k_x$ decreases slightly
as $\beta$ increases.

\subsection{Dependence on Half-Thickness $z_d$}
Fig.~\ref{sig_x_zc} shows the dispersion relation of the case $k_y=0$
for different $z_d$ (the half-thickness of the unperturbed gas disk).
The other parameters take the fiducial values $\alpha=1$, $\beta=1$,
$\kappa_\|=100$ and $\Omega=0$.
Both $\sigma_{x,{\rm max}}$ and $k_{x,{\rm crit}}$ increase as $z_d$ increases.

Fig.~\ref{sig_y_zc} shows the dispersion relation of the case $k_x=0$
for different $z_d$.
The other parameters take the fiducial values $\alpha=1$, $\kappa_\|=100$,
$z_d=1$ and $\Omega=0$.
As $z_d$ increases, $\sigma_{y,{\rm max}}$ gradually increases and
$k_{y,{\rm crit}}$ decreases quite a lot.

Fig.~\ref{sig_xy_zc} shows the dispersion relation of general perturbations
for $z_d=2$ ({\it left}), $0.67$ ({\it middle}), and $0.35$ ({\it right}).
The other parameters take the fiducial values $\alpha=1$, $\beta=1$,
$\kappa_\|=100$ and $\Omega=0$.
When the $z_d$ is large (small) the most unstable mode occurs at $k_y=0$ ($k_x=0$).
When the perturbations develop into elongated fragments, they prefer to align
perpendicular (parallel) to the unperturbed magnetic field.
The transition between parallel and perpendicular fragmentation occurs around
$z_d=0.67$.

\subsection{Dependence on Angular Velocity $\Omega$}
Fig.~\ref{omega} ({\it left}) shows the dependence of the maximum growth rate
on $\beta$ for different $\Omega$ (rotational angular velocity of the disk).
The other parameters take the fiducial values $\alpha=1$, $\beta=1$,
$\kappa_\|=100$ and $z_d=1$.
We note that the maximum growth rate is mostly $\sigma_{x,{\rm max}}$ for the cases
we studied here.
As $\beta$ increases, the maximum growth rate increases slightly
until $\beta\sim 3$ for $\Omega=0.0$, $\beta\sim 5$ for $\Omega=0.5$
and $\beta\sim 9$ for $\Omega=1.0$, then it decreases with $\beta$.
However, in the case of $\Omega=0.0$, the maximum growth rate shifts to 
more or less
constant at $\beta>40$, because the maximum growth rate shifts from
$\sigma_{x,{\rm max}}$ to $\sigma_{y,{\rm max}}$ at $\beta\sim40$.
When we increase $\Omega$, the shift from $\sigma_{x,{\rm max}}$ to 
$\sigma_{y,{\rm max}}$ occurs at larger and larger $\beta$. 
When $\Omega>0.26$ this shift of unstable modes never
substantiates, and the $k_x=0$ mode becomes stabilised 
(see Fig.~\ref{omega}, {\it right}). The growth of perturbations 
in this direction is stabilised by the Coriolis force.
Thus when $\Omega$ is large enough the elongated filaments
always align along the unperturbed magnetic field.

\section{SUMMARY AND DISCUSSION}
This is the first attempt to study how cosmic rays affect the Parker-Jeans
instability of magnetized self-gravitating gaseous disks.
We adopt a two-fluid model for the cosmic-ray-plasma system, in which
the propagation of cosmic rays includes advection and diffusion.
In this work we consider slab geometry with magnetic field lies along the slab,
and cosmic rays diffuse only along the magnetic field lines.
Moreover, we neglect external gravitational force.
We perform standard linear perturbation analysis and work out the cases
of symmetric boundary conditions at $z=0$ (Eq.~(\ref{bc-0}), antisymmetric modes
are less unstable).
In general the most unstable mode occurs at either $k_y=0$ or $k_x=0$,
where the unperturbed initial magnetic field lies in the x-direction.
Thus the perturbations will develop into elongated fragments with
axes either perpendicular to ($k_y=0$) or parallel to ($k_x=0$) the initial
magnetic field.

The coupling between CRs and gas is contained in the diffusion coefficient in
our model.
In the case of $k_y=0$, the growth rate of the instability increases as $\kappa_\|$
increases (i.e., the coupling becomes weaker), and is levelled off at large
$\kappa_\|$ (see Fig.~\ref{sig_x_k}).
Recall that the thermal gas pressure is the stabilizing factor in both Jeans instability
and Parker instability.
One may interpret CR as a gas that can slip through the thermal gas according to
the value of the diffusion coefficient or the strength of the coupling.
During the process it exerts part of its pressure onto the thermal gas.
In effect the thermal gas pressure is augmented and the Parker-Jeans instability
becomes less unstable.
We should point out that the instability cannot be stabilized by just decreasing the
diffusion coefficient (it is less unstable only), except when the diffusion
coefficient is exactly zero and the CRs and thermal gas are tightly bound together.
See Fig.~\ref{sig_x_k} ({\it top}) for detail.
In the case of $k_x=0$, the dispersion relation does not depend on $\kappa_\|$
(see Fig.~\ref{sig_y_k}).
This reflects the assumption that CRs diffuse along the magnetic field lines only.

Keeping other parameters fixed, the increase of $\beta$ (the ratio of unperturbed
CR pressure to gas pressure) has both stabilizing and destabilizing effect
(see Fig.~\ref{betax}).
The wavenumber beyond which the system is stable increases as $\beta$ increases
(it approaches a maximum value when $\beta>3$), i.e., the unstable range increases.
The maximum growth rate increases (slightly) with $\beta$ when $\beta<3$ but
decreases with $\beta$ when $\beta>3$.
We deem that the Parker mode is responsible for the increase in the unstable range and
also for the increase in maximum growth rate when $\beta<3$,
while the Jeans mode is responsible for the decrease in maximum growth rate when
$\beta>3$.
For comparison, Parker instability under constant external gravity (and
no self-gravity is considered) is usually studied in an exponential atmosphere.
As $\beta$ increases, both unstable range and maximum growth rate increase until
some maximum values (see the appendix for a brief discussion).
The effect of CRs on Parker instability under constant external gravity has, of course,
been studied before, for instance, the original work by \citet*{parker66},
and more recent works by \citet*{ryu03} and \citet*{kuwabara04} which have taken the
diffusion of CR into account.
However, one may notice that in 
\citet*{parker66} and \citet*{ryu03},
both unstable range and maximum growth rate
of the instability increase without limit as $\beta$ approaches $\infty$.
We should point out that the apparent increase without bound is just an artefact of
normalisation. In these works the wavenumber and the growth rate are normalised to
$1/H$ and $u/H$, where $H$ is the scale height of the exponential atmosphere which
is proportional to the total unperturbed pressure, and $u$ is a characteristic speed
(say, the isothermal sound speed). Thus the normalisation decreases as $\beta$ increases.
The actual unstable range and maximum growth rate are always finite (see appendix).

\citet*{nagai98} showed that the direction of the longitudinal axis of the elongated
filaments formed by the growth of the perturbation with respect to the magnetic
field depends on the half thickness of the disk or the external pressure.
For thicker (thinner) disk the growth rate of perturbations in the $x$ direction is
larger (smaller) than those in the $y$ direction.
Here we demonstrate that the alignment can also be affected by CR diffusion.
Fig.~\ref{sigxyz} shows the dependence of the growth rate on the half thickness
of the disk $z_d$ and the relation between $\sigma_{x,{\rm max}}$ and $\sigma_{y,{\rm max}}$
for different $\kappa_\|$ in the case of $\alpha=1$ and $\beta=1$.
As long as $\sigma_{x,{\rm max}}<\sigma_{y,{\rm max}}$ the filaments will align along the
magnetic field.
As shown in the figure, when $z_d<0.65$ or $z_d>1.1$,
then $\sigma_{x,{\rm max}}$ is either always smaller or larger than
$\sigma_{y,{\rm max}}$ and the diffusion of CR does not affect the direction of
the filaments.
However, when $0.65<z_d<1.1$ (the grey region in the figure), the alignment depends
on $\kappa_\|$. At a particular $z_d$ the filaments prefer to align along the magnetic
field when $\kappa_\|$ is small enough (i.e., the coupling is strong enough), and
vice versa.

The CR pressure also affect the alignment of filaments.
Fig.~\ref{beta_deps_zc} shows the dependence of the growth rates
$\sigma_{x,{\rm max}}$ and $\sigma_{y,{\rm max}}$ on $\beta$ for three thickness
$z_d=3$, $1$, and $0.66$.
Whenever there is a crossover between the curves $\sigma_{x,{\rm max}}$ and
$\sigma_{y,{\rm max}}$ for the same $z_d$, the direction of the filaments depends on the
CR pressure.
In the case of $z_d=0.66$, $\sigma_{x,{\rm max}}$ is always smaller than $\sigma_{y,{\rm max}}$.
For $z_d=1$ and $z_d=3$, $\sigma_{x,{\rm max}}<\sigma_{y,{\rm max}}$ when
$\beta>40$ and $\beta>550$, respectively (also see Fig.~\ref{w3beta} for the case of $z_d=1$).
In fact, for larger $z_d$ crossover still exists for large enough $\beta$, but the
difference between $\sigma_{x,{\rm max}}$ and $\sigma_{y,{\rm max}}$ is very small in those
range of $\beta$.
Therefore, the alignment of the elongated filament depends on CR pressure when the thickness
of the disk is large enough ($z_d>0.66$).

The dependence of the eigenfunction $\delta\bar\rho$ on several factors
is shown in Fig.~\ref{ef_ro}.
The top panel shows the dependence on the CR diffusion coefficient $\kappa_\|$.
The solid line shows $\delta\bar\rho$ with $\kappa_\|=200$
at the most unstable wave number $(k_x\,,k_y)=(0.5,\,0)$,
and the dotted line shows $\delta\bar\rho$ with any $\kappa_\|<1$
at the most unstable wave number $(k_x\,,k_y)=(0,\,0.35)$.
The middle panel shows the dependence on the half thickness of the disk $z_d$.
The solid line shows the $\delta\bar\rho$ with $z_d=3$
at the most unstable wave number $(k_x\,,k_y)=(0.59,\,0)$
and the dotted line shows $\delta\bar\rho$ with $z_d=0.1$
at the most unstable wave number $(k_x\,,k_y)=(0,\,2.94)$.
The bottom panel shows the dependence on the ratio of CR pressure to gas pressure
$\beta$.
The solid line shows the $\delta\bar\rho$ with $\beta=3$
at the most unstable wave number $(k_x\,,k_y)=(0.51,\,0)$
and the dotted line shows $\delta\bar\rho$ with $\beta=1000$
at the most unstable wave number $(k_x\,,k_y)=(0,\,0.29)$.
The solid line in each panel shows the case when the elongated filaments have the
tendency to align their longitudinal axes perpendicular to the unperturbed magnetic field,
while the dotted line in each panel shows the case when the elongated filaments prefer to align
along the unperturbed magnetic field.
Fig.~\ref{ef_ro} shows that the perturbation of density concentrate around the mid-plane
if the most unstable mode is $k_x=0$ (the dotted line), while it spread more evenly if
the most unstable mode is $k_y=0$ (the solid line).

\acknowledgments
Numerical computations were carried out on the VPP5000 at the National
Astronomical Observatory, Japan (NAOJ) under the projects wtk03b,
and the SX-6 at the National institute of Information and Communications
Technology (NICT), Japan.
C.M. Ko is supported in part by the National Science Council of Taiwan, by grants
NSC-92-2112-M-008-046 and NSC-93-2112-M-008-017.

\appendix

\section{PARKER INSTABILITY IN AN EXPONENTIAL ATMOSPHERE}

In this appendix we briefly describe the dependence of the unstable range and
the maximum growth rate of the Parker instability under constant external gravity
in an exponential atmosphere.

In the unperturbed equilibrium state the external gravity
${\bf g}=-g{\bf e_z}$ and the magnetic field ${\bf B}=B{\bf e_x}$.
The hydrostatic equilibrium equation ($g$ is constant)
\begin{equation}\label{hydrostatic}
  {\dd P_{\rm t}\over\dd z}=-\rho g\,,
\end{equation}
gives an exponential atmosphere if $P_{\rm t}\propto\rho$.
Specifically, $P_{\rm t}=P_{\rm t0}\exp[-z/H]$ and $\rho=\rho_0\exp[-z/H]$.
The scale height $H=P_{\rm t}/(\rho g)=P_{\rm t0}/(\rho_0 g)$ is a constant.
The total pressure $P_{\rm t}$ includes thermal gas pressure, magnetic pressure
and other pressures such as cosmic ray pressure.

To study the linear stability of the system, one may perform standard normal mode analysis.
We present the result of a particular simple case: only gas and magnetic field are actively
involved in the dynamics of the system. The other components, e.g., cosmic rays, just
provide the background pressure against the external gravity in the unperturbed equilibrium
state. This is the case when cosmic ray diffusion coefficient is very large, i.e.,
the coupling is vanishing small.
Let the perturbations be proportional to $\exp(\sigma t+i k_xx)$
(forget about perturbations in y-direction to further simplify discussion).
The dispersion relation of the lowest mode in z-direction is
\begin{equation}\label{dispersion}
  \sigma^4+\left(V^2_{\rm A}+C^2_{\rm s}\right)\left(k_x^2+{1\over 4H^2}\right)\sigma^2
  +k_x^2C^2_{\rm s}\left[\left(k_x^2+{1\over 4H^2}\right)V^2_{\rm A}+N^2\right]=0\,,
\end{equation}
where $V_{\rm A}=B/\sqrt{\mu_0\rho\,}$ and $C_{\rm s}=\sqrt{\gamma_{\rm g} P_{\rm g}/\rho\,}$
are the Alfv\'en velocity and adiabatic sound speed, respectively;
$N=\sqrt{g/H-g^2/C^2_{\rm s}\,}$ is the Brunt-V\"ais\"al\"aa frequency (with contributions
from magnetic field and cosmic rays through $H$).
The system is susceptible to Parker instability if
\begin{equation}\label{criterion}
  N^2+{V^2_{\rm A}\over 4 H^2}
  ={g\over H}-{g^2\over C^2_{\rm s}}+{V^2_{\rm A}\over 4 H^2}
  <0\,.
\end{equation}

To facilitate discussion, we suppose $P_{\rm B}=B^2/2\mu_0=\alpha P_{\rm g}$,
$P_{\rm c}=\beta P_{\rm g}$ and $P_{\rm g}=\rho u^2$ (the isothermal sound speed $u$ is
constant). Thus $P_{\rm t}=P_{\rm g}+P_{\rm B}+P_{\rm c}=(1+\alpha+\beta)P_{\rm g}
=(1+\alpha+\beta)\rho u^2$.
A natural time scale of the system is $u/g$ and a natural length scale is $u^2/g$.
In these units the parameters in Eq.~(\ref{dispersion}) become:
$V^2_{\rm A}=2\alpha$, $C^2_{\rm s}=\gamma_{\rm g}$, $H=(1+\alpha+\beta)$ and
$N^2=-(1+\alpha+\beta-\gamma_{\rm g})/(1+\alpha+\beta)/\gamma_{\rm g}$.
The growth rate and wavenumber become $\sigma u/g$ and $k_x u^2/g$.
The instability criterion Eq.~(\ref{criterion}) can then be written as
\citep[cf.,][]{parker66}
\begin{equation}\label{criterion-n}
  (1+\alpha+\beta)^2>\gamma_{\rm g}\left(1+{\textstyle{3\over 2}}\alpha+\beta\right)\,.
\end{equation}
In these units it can readily be shown that the critical wavenumber beyond which the
system is stable is given by
\begin{equation}\label{kcrit}
  \left(k_{x,{\rm crit}}{u^2\over g}\right)^2
  ={1\over 2\alpha\gamma_{\rm g}}
  \left[1-{\gamma_{\rm g}\left(1+{\textstyle{3\over 2}}\alpha+\beta\right)
  \over (1+\alpha+\beta)^2}\right]\,,
\end{equation}
and the maximum growth rate is given by the smaller root of
\begin{eqnarray}\label{sigmamax}
  (2\alpha-\gamma_{\rm g})\left(\sigma_{x,{\rm max}}{u\over g}\right)^4
  -2(2\alpha+\gamma_{\rm g})
  \left[1-{\gamma_{\rm g}\left(1+{\textstyle{1\over 2}}\alpha+\beta\right)
  \over (1+\alpha+\beta)^2}\right]\left(\sigma_{x,{\rm max}}{u\over g}\right)^2 \nonumber \\
  +\left[1-{\gamma_{\rm g}\left(1+{\textstyle{3\over 2}}\alpha+\beta\right)
  \over (1+\alpha+\beta)^2}\right]^2=0\,.
\end{eqnarray}
In fact, $(k_{x,{\rm crit}}u^2/g)^2$ and $(\sigma_{x,{\rm max}}u/g)^2$ increase monotonically
with $\beta$.
Moreover, as $\beta\rightarrow\infty$, 
$(k_{x,{\rm crit}}u^2/g)^2\rightarrow 1/2\alpha\gamma_{\rm g}$
and $(\sigma_{x,{\rm max}}u/g)^2\rightarrow 1/(\sqrt{2\alpha}+\sqrt{\gamma_{\rm g}})$.

\clearpage


\begin{figure}
\caption{{\it Left}: Gas pressure in the unperturbed equilibrium state 
                     as a function of $z$
                     (vertical coordinate) in the case of $\alpha=1$ 
                     and $\beta=1$.
         {\it Right}: Gravity in the unperturbed equilibrium state 
                      as a function of $z$.}
\label{initial_dist}
\end{figure}


\begin{figure}
\caption{{\it Top}: The dependence of the dispersion relation at 
                    $k_y=0$ on $\kappa_\|$,
                    the CR diffusion coefficient.
                    $\sigma_x$ is the growth rate when $k_y=0$ and 
                    $k_x$ is the wavenumber
                    in the direction of the unperturbed magnetic field.
                    The other parameters are $\alpha=1$, $\beta=1$, $z_d=1$ 
                    and $\Omega=0$.
         {\it Bottom}: The dependence of the maximum growth rate 
                       $\sigma_{x,{\rm max}}$ on $\kappa_\|$.}
\label{sig_x_k}
\end{figure}


\begin{figure}
\caption{The dependence of the dispersion relation at $k_x=0$ on $\kappa_\|$.
         $\sigma_y$ is the growth rate when $k_x=0$.
         The other parameters are the same as in Fig.~\ref{sig_x_k}.}
\label{sig_y_k}
\end{figure}


\begin{figure}
\caption{The general dispersion relation for the Parker-Jeans instability
         against the effect of CR diffusion.
         {\it Left}: $\kappa_\|=100$,
         {\it Middle}: $\kappa_\|=1$, and
         {\it Right}: $\kappa_\|=0.1$.
         The other parameters are the same as in Fig.~\ref{sig_x_k}.}
\label{sig_xy_k}
\end{figure}


\begin{figure}
\caption{The dependence of the dispersion relation at $k_y=0$ on $\beta$,
         the ratio of the unperturbed CR pressure to gas pressure.
         $\sigma_x$ is the growth rate when $k_y=0$.
         The other parameters are $\alpha=1$, $\kappa_\|=100$, $z_d=1$ 
         and $\Omega=0$.}
\label{betax}
\end{figure}


\begin{figure}
\caption{The dependence of the dispersion relation at $k_x=0$ on $\beta$.
         $\sigma_y$ is the growth rate when $k_x=0$.
         The other parameters are the same as in Fig.~\ref{betax}.}
\label{betay}
\end{figure}


\begin{figure}
\caption{The general dispersion relation for the Parker-Jeans instability
         against the effect of CR pressure.
         {\it Left}: $\beta=1$,
         {\it Middle}: $\beta=40$, and
         {\it Right}: $\beta=100$.
         The other parameters are the same as in Fig.~\ref{betax}.}
\label{w3beta}
\end{figure}


\begin{figure}
\caption{The dependence of the dispersion relation at $k_y=0$ on $z_d$,
         the half-thickness of the unperturbed gas disk 
         (can be viewed as a measure of external pressure).
         $\sigma_x$ is the growth rate when $k_y=0$.
         The other parameters are $\alpha=1$, $\beta=1$, $\kappa_\|=100$ 
         and $\Omega=0$.}
\label{sig_x_zc}
\end{figure}


\begin{figure}
\caption{The dependence of the dispersion relation at $k_x=0$ on $z_d$.
         $\sigma_y$ is the growth rate when $k_x=0$.
         The other parameters are the same as in Fig.~\ref{sig_x_zc}.}
\label{sig_y_zc}
\end{figure}


\begin{figure}
\caption{The general dispersion relation for the Parker-Jeans instability
         against the effect of the half-thickness of the gas disk.
         {\it Left}: $z_d=2$,
         {\it Middle}: $z_d=0.67$, and
         {\it Right}: $z_d=0.35$.
         The other parameters are the same as in Fig.~\ref{sig_x_zc}.}
\label{sig_xy_zc}
\end{figure}


\begin{figure}
\caption{{\it Left}: The dependence of the maximum growth rate 
                     $\sigma_{{\rm max}}$
                     on $\beta$ (the ratio of unperturbed CR pressure to 
                     gas pressure)
                     at different angular velocity of the gas disk $\Omega$.
                     The other parameters are $\alpha=1$, $\beta=1$, 
                     $\kappa_\|=100$ and $z_d=1$.
         {\it Right}: The general dispersion relation in the case of
                     $\alpha=1$, $\beta=1$, $\kappa_\|=100$, $z_d=1$ 
                     and $\Omega=0.5$}.
\label{omega}
\end{figure}


\begin{figure}
\caption{The dependence of the maximum growth rate on the half-thickness $z_d$.
         $\sigma_{x,{\rm max}}$ is the maximum growth rate at $k_y=0$ at a given
         $\kappa_\|$.
         $\sigma_{y,{\rm max}}$ is the maximum growth rate at $k_x=0$
         ($\sigma_{y,{\rm max}}$ is independent of $\kappa_\|$, 
         see Fig.~\ref{sig_y_k}).
         The other parameters are $\alpha=1$, $\beta=1$ and $\Omega=0$.
         In the grey region, CR diffusion determines which mode is more 
         unstable, thus
         determines the alignment of the perturbations with respect to the 
         unperturbed magnetic field.}
\label{sigxyz}
\end{figure}


\begin{figure}
\caption{The dependence of the maximum growth rate on the ratio of 
         CR pressure to gas pressure $\beta$ at different $z_d$.
         The other parameters are $\alpha=1$, $\kappa_\|=100$ and $\Omega=0$.}
\label{beta_deps_zc}
\end{figure}


\begin{figure}
\caption{The eigenfunction $\delta\bar{\rho}$ of the most unstable mode.
         {\it Top}: The dependence on the CR diffusion coefficient $\kappa_\|$.
                    The solid line is the case of $\kappa_\|=200$ at the 
                    wavenumber
                    $(k_x\,,k_y)=(0.5,\,0)$, and the dotted line is the case of
                    any $\kappa_\|<1$ at the wavenumber 
                    $(k_x\,,k_y)=(0,\,0.35)$.
                    The other parameters are $\alpha=1$, $\beta=1$, $z_d=1$ 
                    and $\Omega=0$.
         {\it Middle}: The dependence on the half-thickness $z_d$.
                    The solid line is the case of $z_d=3$ at 
                    $(k_x\,,k_y)=(0.59,\,0)$,
                    and the dotted line is the of $z_d=0.1$ at 
                    $(k_x\,,k_y)=(0,\,2.94)$.
                    The other parameters are $\alpha=1$, $\beta=1$, 
                    $\kappa_\|=100$ and $\Omega=0$.
         {\it Bottom}: The dependence on the ratio of CR pressure to gas 
                    pressure $\beta$.
                    The solid line is the case of $\beta=3$ at 
                    $(k_x\,,k_y)=(0.51,\,0)$,
                    and the dotted line shows the case of $\beta=1000$ at 
                    $(k_x\,,k_y)=(0,\,0.29)$.
                    The other parameters are $\alpha=1$, $\kappa_\|=100$, 
                    $z_d=1$ and $\Omega=0$.}
\label{ef_ro}
\end{figure}







\begin{thebibliography}{}
\bibitem[Brezzinskii et al.(1990)]{berezinskii90} Berezinskii, V. S.,
     Bulanov, S. V., Dogiel, V. A., Ginzburg, V. L., \& Ptuskin, V. S.
     1990, Astrophysics of Cosmic Rays,
     ed. V. S. Verezinskii \& V. L. Ginzburg (New York: North-Holland)
\bibitem[Chou et al.(2000)]{chou00} Chou, W., Matsumoto, R.,
     Tajima, T., Umekawa, M., \& Shibata, K. 2000,\apj, 538, 710
\bibitem[Drury(1983)]{drury83} Drury, L.O'C. 1983, Rep. Prog. Phys., 46, 973
\bibitem[Drury \& V\"olk(1981)]{drury81} Drury, L.O'C., \& V\"olk, H. J.
     1981, \apj, 248, 344
\bibitem[Elmegreen \& Elmegreen(1978)]{elmegreen78} Elmegreen, B. G.,
      \& Elmegreen, D. M. 1978, \apj, 220, 1051
\bibitem[Elmegreen(1982a)]{elmegreen82a} Elmegreen, B. G. 1982a, \apj, 253, 634
\bibitem[Elmegreen(1982b)]{elmegreen82b} Elmegreen, B. G. 1982b, \apj, 253, 655
\bibitem[Ferri\`ere(2001)]{ferriere01} Ferri\`ere, K. M. 2001, RMP, 73, 1031
\bibitem[Gaisser(1990)]{gaisser90} Gaisser, T. K. 1990, Cosmic Rays and Particle Physics
      (Cambridge: Cambrigde University Press)
\bibitem[Giacalone \& Jokipii(1999)]{giacalone99} Giacalone, J., \& Jokipii, J. R. 1999,
     \apj, 520, 204
\bibitem[Hanasz(1997)]{hanasz97II} Hanasz, M. 1997,
     \aap, 327, 813
\bibitem[Hanasz \& Lesch(1997)]{hanasz97I} Hanasz, M., \& Lesch, H. 1997,
     \aap, 321, 1007
\bibitem[Hanasz \& Lesch(2000)]{hanasz00} Hanasz, M., \& Lesch, H. 2000,
     \apj, 543, 235
\bibitem[Hanasz \& Lesch(2003)]{hanasz03} Hanasz, M., \& Lesch, H. 2003,
     \aap, 412, 331
\bibitem[Hanasz et al.(2004a)]{hanasz04I} Hanasz, M.,
      Kosinski, R., \& Lesch, H. 2004, \apss, 289, 303
\bibitem[Hanasz et al.(2004b)]{hanasz04II} Hanasz, M., Kowal, G.,
      Otmianowska-Mazur, K., \& Lesch, H. 2004, \apj, 605, L33
\bibitem[Hanawa et al.(1992)]{hanawa92} Hanawa, T., Nakamura, F.,
      \& Nakano, T. 1992, \pasj, 44, 509
\bibitem[Kim et al.(1997)]{kim97} Kim, J., Hong, S. S., \& Ryu, D. 1997,
      \apj, 485, 228
\bibitem[Kim \& Hong(1998)]{kim98} Kim, J., Hong, S. S. 1998,
      \apj, 507, 254
\bibitem[Kim et al.(2002)]{kim02} Kim, W. T., Ostriker, E. C.,
    \& Stone J. M. 2002
\bibitem[Ko(1992)]{ko92} Ko, C. M. 1992, \aa, 259, 377
\bibitem[Kuwabara et al.(2004)]{kuwabara04}
     Kuwabara, T., Nakamura, K., \& Ko, C. M., \apj, 607, 828
\bibitem[Lee \& Hong(2005)]{lee05} Lee, S. M., \& Hong, S. S. 2005,
    Arxiv, http://arxiv.org/abs/astro-ph/0503013
\bibitem[Lee et al.(2001)]{lee01} Lee, S. M., Hong, S. S., \&
    \& Kim, J. 2001, JKAS, 34, L285
\bibitem[Nagai et al.(1998)]{nagai98} Nagai, T.,
    Inutsuka, S., \& Miyama, M. 1998, \apj, 506, 306
\bibitem[Parker(1966)]{parker66} Paker, E. N. 1996, \apj, 145, 811
\bibitem[Ptuskin(2001)]{ptuskin01} Ptuskin, V. S. 2001, Space Sci. Rev.,
    99, 281
\bibitem[Ryu et al.(2003)]{ryu03} Ryu, D., Kim, J., Hong, S. S.,
     \& Jones, T. W. 2003, \apj, 589, 338


\end{thebibliography}
\end{document}